\documentclass{article}
\usepackage[a4paper, total={7.2in, 9.5in}]{geometry}
\usepackage[utf8]{inputenc}
\usepackage{blindtext}
\usepackage{ragged2e}
\usepackage{multicol}
\usepackage[font=sf]{caption}
\usepackage{fancyhdr}
\usepackage[pdftex]{hyperref}
\usepackage[dvipsnames]{xcolor}
\usepackage[most]{tcolorbox}
\usepackage{tabularx}
\usepackage{enumitem,kantlipsum}
\usepackage[font=sf]{caption}
\usepackage[font=sf]{floatrow}

\definecolor{textgreen}{RGB}{33, 143, 38}
\definecolor{headergreen}{RGB}{19, 88, 97}

\newcommand\sqbullet{\fcolorbox{black}{textgreen}{\rule{0pt}{3pt}\rule{3pt}{0pt}}}

\begin{document}
\sffamily

\noindent{\Huge Accessibility in astronomy for the visually impaired}

\vspace{0.2cm}{\large{\color{textgreen}\noindent Jake Noel-Storr$^{1,\star}$ and Michelle Willebrands$^{2}$}}\\
{\em $^{1}$Faculty of Science and Engineering and Kapteyn Astronomical Institute, University of Groningen, Groningen, The Netherlands. $^{2}$Leiden Observatory, Universiteit Leiden, Leiden, The Netherlands.}\\
\noindent Email: $^{\star}$\href{mailto:j.noel-storr@rug.nl}{j.noel-storr@rug.nl}
 
\vspace{0.3cm}\noindent{\large We spoke with four researchers to understand the accessibility challenges in astronomy research, education and outreach for blind and visually impaired (BVI) persons, as well as solutions to these challenges and how it innovates data analysis methods for all astronomers.}

\vspace{0.3cm}{
\noindent Those interviewed: \textbf{Nicolas Bonne (NB)} from the Institute of Cosmology \& Gravitation, University of Portsmouth, UK.
\noindent\textbf{Cheryl Fogle-Hatch (CFH)} from Museum Senses LLC, Baltimore, Maryland, USA.
\noindent\textbf{Garry Foran (GF)} from the Centre for Astrophysics and Supercomputing, Swinburne University of Technology, Australia.
\noindent\textbf{Enrique Perez Montero (EPM)} from the Instituto de Astrofísica de Andalucía, Granada, Spain.

}

\begin{multicols*}{3}
\noindent$\sqbullet{}$ \textbf{Question: Could you please introduce yourselves and tell us how you got involved with accessibility in astronomy?}

\noindent\textbf{NB: }
I have a severe vision impairment but chased the dream of beginning a career in astronomy nonetheless. I started an astronomy PhD programme but dropped out because it was too challenging due to my impairment. After starting again some years later, this time choosing a project that was a bit less visually demanding, I learned to work around some of the challenges. This got me thinking about wider issues of accessibility and I started working on this too.

\noindent\textbf{CFH: }
I am an archaeologist by training, studying stone spear points in America. My current occupation is as an independent consultant. I have a lot of interest in tactile and sound. At a conference a couple of years ago, I got interested in accessibility in astronomy. I have a desire to have a tool that could load any sort of data, from any subject, and investigate it.

\noindent\textbf{GF: }
After losing my sight to a degenerative retinal condition, I retired from my position as a research scientist working in the field of synchrotron radiation technology and X-ray physics.  Still keen to pursue my passion for discovery, I recently completed my PhD in the Centre for Astrophysics and Supercomputing at Swinburne University of Technology. Relying entirely on audio feedback to carry out my research work, I am active in a collaboration that is developing sonification tools that facilitate the management and analysis of astrophysical data using sound.  

\noindent\textbf{EPM: }
I am an astrophysicist, and senior scientist in the Instituto de Astrofísica de Andalucía. My research is related to studying how galaxies evolve. I was diagnosed 25 years ago, when graduating from my undergraduate years and starting my PhD, while still having full vision. My eyesight has degenerated slowly over time and I am now completely blind. I have started an outreach programme with a key message: you can do astronomy without seeing.

\vspace{0.1cm}\noindent {\color{textgreen} {\em \textbf{\em Commentary:} Workshop participants were interested in the current state of the field in research, education, and outreach in astronomy from the perspective of visually impaired individuals. While we are commonly aware of the adage that “astronomy is a highly visual science” how does this translate into a barrier for the participation of the blind and visually impaired. We posed this to our panel.}}

\vspace{0.2cm}\noindent$\sqbullet{}$ \textbf{Question: Why is the way we traditionally practice astronomy research, education and outreach inaccessible for the blind and visually impaired?}

\noindent\textbf{NB: } 
Traditionally, you have a picture of an astronomer with their eye up against a telescope. But we don’t really do this anymore, we use instruments to detect light on the back of telescopes which is then recorded digitally. This is reflected in the way astronomy is presented to the public. We lean very heavily on inspiring astronomy images in presenting astronomy. We also rely heavily on visual descriptions which are not so accessible. For example, talking about colour and not why something has its colour. For someone with vision, 80\% of information is perceived through their eyes so it is natural that we would communicate this way, but it is not accessible to BVI persons.

\noindent\textbf{CFH: }
I think that it is lazy teaching. The visual aspect is expected in astronomy and it is a case of ‘we have always done it that way’. When I talk to students of all levels of sight they say that they benefit from multi-sensory teaching. The education system is stuck in the visual output. Relying on writing on the whiteboard and using pictures when we could use tactile approaches, for example. And when it comes to research, there are just times that you need to get things described and have to rely on a colleague. Audio guides and descriptions might be okay but you are relying on the filter of somebody else, lacking the opportunity to make your own discovery and exploration. You are relying on somebody else's perception.

\noindent\textbf{EPM: }
One of the problems for new people entering the field is realising that it is possible to be an astronomer with a visual impairment. We need to raise awareness that it is possible to be BVI and carry out science. If teachers and people in universities know that it is possible to have accessible resources, everyone will benefit from that. If BVI persons want to be devoted to science, they can do it. 

\noindent\textbf{GF: }
When I wanted to pursue a career in astronomy, it wasn’t obvious by any means that it was going to be possible. It is a very visual field, both to professional and lay observers - when you think about astronomy and astrophysics you think about awe-inspiring images. So awareness is one thing. Being ready to struggle is also important. You need a broader way of thinking and a collaborative way of doing research. For someone who is visually impaired, knowing that you can work with others within a team and rely on colleagues has been important to enabling my progress in the field.. 

\vspace{0.1cm}\noindent {\color{textgreen} {\em \textbf{\em Commentary:} We can see from the responses that our perspectives on astronomical data could, and maybe should, change to be more inclusive. We tend to translate all data into images, as an historical vestige of the days of gazing through telescopes on mountaintops. Other ways of interpreting data using other senses could indeed improve accessibility and learning for everyone, not just the blind and visually impaired. Pursuing this idea, we went on to ask the following.}}

\vspace{0.2cm}\noindent$\sqbullet{}$ \textbf{Question: How can we, as a community, make astronomy more BVI accessible? Both in research and in education, to both attract more students to astronomy and STEM fields and also allow for diversely abled people to become astronomers.}

\noindent\textbf{NB: }
A multi-modal approach is probably always the best way to go. As many sensory modes as possible will be the best way forward. For example, with our tactile resources, we include the visual element as well. 

\noindent\textbf{EPM: }
I use sound in outreach and dissemination in general, even in seminars with my colleagues. Sonification of astronomical data and using sound in presentations to present results benefit everyone in the audience. However, sounds without context are hard to interpret. If you play a sonification of astronomical data without any context it is meaningless. This context can be given in an article or voice description. In addition, using the context of touch is very important in outreach. For example, tactile models of the sky map of the northern sky.

\noindent\textbf{GF: }
In terms of the adaptive technologies that are available, one of the biggest challenges is that there are almost no two people whose needs are the same.  Therefore, it is very difficult to develop a single set of tools that works for everyone. My work with my colleagues using sound has shown me that there can be a way through.  Most people are quite surprised when they hear I work in astrophysics and yet have no vision. The digital age has given us plenty of scope to improve the accessibility of the field. There are a few things that immediately come to mind.  Something mundane is the accessibility of materials in journal articles.There has been some progress, but there are still challenges in mathematical and graphical content. Sonification has potential to address some of these issues, and in some contexts has potential advantages over vision alone. There is, after all, a reason that we have ears as well as eyes! We can also think about touch. A multi-modal approach, with a combination of sound and haptic feedback, would be a great way forward.

\noindent\textbf{NB: } 
Besides, for astronomy researchers, the sonification might not need to sound nice but needs to be useful. For the public and in education, the emotional response might be more important. If the sound scares people or is unpleasant it might put them off. In general, we need to demonstrate through publications that sonification and other senses are successful and valid, and there are unique benefits to multi-sensory approaches. If there is a standard framework it might be easier to get people involved and introduce them to this accessible methodology. 

\vspace{0.1cm}\noindent {\color{textgreen} {\em \textbf{\em Commentary:} Commentary: It is clear from this that there are many ways to make astronomy research and education more inclusive and accessible. Even though there are still some challenges that need to be addressed, the work being done by this community also leads to new perspectives and possibilities of practicing astronomy that could potentially benefit all astronomers.}}

\vspace{0.2cm}\noindent$\sqbullet{}$ \textbf{Question: Have these new technologies/methods for the visually impaired led to innovative ways of doing research or education that is useful for non-BVI people as well?}

\noindent\textbf{GF: }
The answer to that is a resounding yes. It is well known that improved accessibility actually helps everyone in the end. Making tools such as software more accessible makes the experience even richer for someone with full sensory capacity. While the primary motivation for someone like myself is making it possible for BVI persons to do astronomy research, there is no question that there is an opportunity for a broader scope of applicability. Our ears, for example, have tremendous dynamic range and temporal resolution (not a great spatial resolution, but great spatial coverage). Sonification enables us to distinguish very faint sounds in a noisy background, which is known as the 'Cocktail party' effect.

\noindent\textbf{NB: }
Our ears are very sensitive and we often don’t use this enough. For example, it is amazing what sounds we can pick out from the context of a noisy room. I can say that it is amazing what signal you can pick out from noisy data in sound - which might be harder to see in a noisy plot. In sound, we are good at picking out patterns as well. What a lot of this is going to take though is a cultural shift. A lot of astronomers see this as a gimmick and not something with practical applications. There is work to be done to make sonification more mainstream and show that it is a valid way of analysing data.

\noindent\textbf{CFH: }
I can’t tell you the number of times I’ve been walking with someone sighted and they haven’t heard or smelt something in their environment, for example, heard a fountain or smelled a coffee shop. People need to pay more attention to their other senses. Raising awareness of these benefits will be important to mainstream accessible approaches in research and education. 

\noindent\textbf{EPM: } 
Audio description has incredible success amongst all types of people. When you are describing what is in the image clearly, it benefits everyone to get the message. 

\vspace{0.1cm}\noindent {\color{textgreen} {\em \textbf{\em Commentary:} The innovations in practicing astronomical research and education are exciting and have the potential to benefit the wider community and not only BVI persons. However, progress is still necessary to improve the accessibility of the field. So, to close off the interview, we asked each speaker:}}

\vspace{0.2cm}\noindent$\sqbullet{}$ \textbf{Question: What do you think is the biggest challenge that we still need to address for BVI people in astronomy?}

\noindent\textbf{GF: }
It is hard to generalise and I do not speak for everyone. But the graphical presentation of data, like plots and figures, remains as a big barrier. The other is that there needs to be a bottom-up approach to include accessibility in the design of new facilities and tools. I would like to think I can re-engage with my interests in instrumentation and spectroscopy, but at the moment the interaction is completely inaccessible. When designing new tools, people need to adopt a user-centred approach: for example, how might somebody who can’t see a graphical user interface do these things?  I come from an era when everything could be done from a command line on a computer terminal , which was accessible to me. Whilst GUIs are nice, accessible interfaces and approaches could still be considered.

\noindent\textbf{NB: } 
Being able to access and interpret data ourselves is very important so that you are not relying on other people's perceptions and interpretations. If we are just presenting somebody with a final product we are missing something. 

\noindent\textbf{EPM: }
Mainly the plots are a challenge for me. Recently I received a request to review an article, and, as I am forced to do since I totally lost my vision, I have had to say no.

\noindent\textbf{CFH: }
Getting people to accept and be aware of the importance of accessibility and therefore research it. So that ultimately, the community will change their practices and become more inclusive.

\noindent\textbf{NB: }
It is hard to get people to change their practices though. Accessibility approaches need to be made more mainstream early on in the education system. 

\vspace{0.1cm}\noindent{\color{textgreen} {\em \textbf{\em Commentary:} We have heard many ways in which we can make astronomy more accessible to the visually impaired, and also how those things extend much more widely across our community. We maybe all use sloppy language such as “as you can see from this plot” when really we should understand that this language is excluding both the BVI community, but also those with good visual acuity… who just also don’t know what you are talking about. Giving descriptions of meaning and sharing data in different ways opens access to everyone.}}
 
\vspace{0.3cm}\noindent{\color{textgreen}Competing interests}

\noindent{\small The authors declare no competing interests.}

\end{multicols*}

\end{document}